\documentclass[12pt]{iopart}
\usepackage{iopams} 
\usepackage{graphicx}
\usepackage{color}
\begin{document}

\title[]{Polaron in the dilute critical Bose condensate}

\author{Volodymyr~Pastukhov}

\address{Department for Theoretical Physics, Ivan Franko National University of Lviv,\\ 12 Drahomanov Street, Lviv-5, 79005, Ukraine}
\ead{volodyapastukhov@gmail.com}
\vspace{10pt}
\begin{indented}
\item[] 
\end{indented}

\begin{abstract}
The properties of impurity immersed in the dilute $D$-dimensional Bose gas at temperatures close to the second-order phase transition point are considered. Particularly by means of the $1/N$-expansion we calculated the leading-order polaron energy and the damping rate in the limit of vanishing boson-boson interaction. It is show that the perturbative effective mass and the quasiparticle residue diverge logarithmically in the long-length limit signalling the non-analytic behavior of impurity spectrum and a non-pole structure of a polaron Green's function in the infrared region, respectively.
\end{abstract}


\vspace{2pc}

\section{Introduction}

Recently the Bose polaron problem attracted much attention of researches both experimentalists \cite{Schmid_et_al,Spethmann_et_al,Hu,Jorgenzen} and theorists \cite{Novikov_Ovchinnikov_1,Novikov_Ovchinnikov_2,Rath,Li,Christensen,Grusdt_et_al,Vlietinck_et_al,Ardila_1,Ardila_2,Panochko_1,GSSD,Panochko_2}. Although the most characteristic properties of a single impurity immersed in the superfluid were studied in the context of a dilute $^3$He-$^4$He solution \cite{Landau_Pomeranchuk} in late forties, the problem is still of current interest. Moreover, the experimental possibility to fine-tune a value and even a sign of the boson-impurity coupling parameter stimulated further research and allowed to predict numerous phenomena including localization-delocalization transition \cite{Cucchietti,Roberts}, non-typical dynamics \cite{Shchadilova,Lampo} and thermalisation \cite{Lausch} of the moving Bose polaron, emergence of the Efimov physics \cite{Zinner,Levinsen,Naidon,Sun,Yoshida}, etc. The low-dimensional \cite{Catani_et_al,Burovski,Parisi,Grusdt,Volosniev,Pastukhov_1D} and the mixed-dimensional \cite{Loft} systems also possess very non-trivial behavior. However, less is known about the finite-temperature phase diagram of the Bose polarons, especially in a narrow region of the Bose-Einstein condensation (BEC) transition point, where only the equal-mass limit \cite{Guenther,Levinsen_et_al} is studied. Particularly in Ref.~\cite{Levinsen_et_al} the perturbative second-order energy correction and the damping were found to increase enormously near the BEC temperature. On the other hand, from the theory of the critical Bose systems it is known \cite{VPP_2} that the reliable result for the one-particle spectrum can be obtained only with an infinite series of diagrams taken into account. In the present article by means of the $1/N$-expansion, which effectively incorporates bosonic ``particle-hole'' bubbles, we address to the problem of an impurity weakly coupled to the dilute $D$-dimensional ($D>2$) bath that undergoes the BEC transition.

\section{Model and method}
The possibility to study the properties of a single impurity atom immersed in the Bose condensate as a Fermi-Bose mixture with vanishing density of fermions opens up a new opportunities for the use of field-theoretical methods to this problem.
For instance, adopting path-integral formulation one readily writes down the Euclidean action of our model
\begin{eqnarray}\label{A}
	\mathcal{A}=\mathcal{A}_0+\mathcal{A}_B+\mathcal{A}_{int},
\end{eqnarray}
where the first term describes ideal Fermi gas with dispersion $\varepsilon_f(p)=\hbar^2p^2/2m_f$
\begin{eqnarray}\label{A_0}
	\mathcal{A}_0=\sum_{P}\{i\nu_p-\varepsilon_f(p)+\mu_f\}\psi^*_{P}
	\psi_{P},
\end{eqnarray}
the second one
\begin{eqnarray}\label{A_B}
	\mathcal{A}_B=\sum_{K}\{i\omega_k-\varepsilon_{k}+\mu\}\phi^*_{\sigma K}\phi_{\sigma K}
	-\frac{gT}{2NV}\sum_{K,Q,S}\phi^{*}_{\sigma Q}\phi^{*}_{\sigma' S}\phi_{\sigma'S-K}\phi_{\sigma Q+K},
\end{eqnarray}
is the action of the $N$-component Bose system ($\varepsilon_k=\hbar^2k^2/2m$, $V$, $T$  are a volume and a temperature of the system respectively and summations over indices $\sigma,\sigma'=1,2\ldots, N$ are understood). The third term $\mathcal{A}_{int}$ of action (\ref{A}) takes into account interaction of the impurity atom with bosonic bath 
\begin{eqnarray}\label{A_int}
	\mathcal{A}_{int}=-\frac{\tilde{g}T}{NV}\sum_{K,Q,P}\phi^{*}_{\sigma Q}\phi_{\sigma Q+K}\psi^{*}_{P}\psi_{P-K}.
\end{eqnarray}
Let us stress once again that the introduction of $N$-component model of the environment is a formal tool which helps to classify diagrams in further calculations and only the $N=1$ choice leads to the physically relevant case, while results are applicable for $N\gg 1$. The presented here chemical potentials $\mu$ and $\mu_f$ control the number of Bose and Fermi particles, respectively. Other notations are also typical: the arguments of fermionic Grassmann  $\psi^*_{P}$, $\psi_{P}$ and bosonic complex $\phi^*_{\sigma K}$, $\phi_{\sigma K}$ fields are the $D+1$ ``wave-vectors'' $P=(\nu_p,{\bf p})$ and $K=(\omega_k,{\bf k})$ with $\nu_p$ and $\omega_k$ standing for odd (fermionic) and even (bosonic) Matsubara frequencies. The coupling constants controlling the interaction strength can be related to the $s$-wave scattering lengths in $D$ dimensions 
\begin{eqnarray}
	g=\frac{4\pi^{D/2}\hbar^2 a^{D-2}}{\Gamma(D/2-1)m}, \ \ \ \ \tilde{g}=\frac{2\pi^{D/2}\hbar^2 \tilde{a}^{D-2}(m+m_f)}{\Gamma(D/2-1)mm_f}.
\end{eqnarray}
Introducing auxiliary field $\varphi_K$ that decouples by means of the Hubbard-Stratonovich transformation quartic term in Eq.~(\ref{A_B}) and making use of change of variables in the path integral $\varphi_K=\varphi'_K+\frac{i\tilde{g}\sqrt{T}}{N\sqrt{V}}\sum_P\psi^{*}_{P}\psi_{P+K}$
we finally rewrite action (\ref{A}) as follows
\begin{eqnarray}\label{A_eff}
	\mathcal{A}=\sum_{K}\{i\omega_k-\varepsilon_{k}+\tilde{\mu}\}\phi^*_{\sigma K}\phi_{\sigma K} +\sum_{P}\{i\nu_p-\varepsilon_f(p)+\tilde{\mu}_f\}\psi^*_{P}
	\psi_{P}
\nonumber\\
		-\frac{N}{2g}\sum_K\varphi_K \varphi_{-K}-i\sqrt{\frac{T}{V}}\sum_{K,Q}\varphi_K\phi^{*}_{\sigma Q}\phi_{\sigma Q-K}-i\frac{\tilde{g}}{g}\sqrt{\frac{T}{V}}\sum_{K,P}\varphi_K\psi^{*}_{P}\psi_{P+K},
\end{eqnarray}
where prime near field $\varphi_K$ is dropped, $\varphi_K|_{{\bf k}= 0}=0$ and $\tilde{\mu}=\mu-ng$, $\tilde{\mu}_f=\mu_f-n\tilde{g}$ ($n$ is the density of each boson component) denote the Hartree-shifted chemical potentials. After all transformations Eq.~(\ref{A_eff}) constitutes the action of mixture of ideal Bose and Fermi gases coupled to the ``phonon'' field $\varphi_K$. In such a formulation the properties of the impurity atom are fixed by this collective mode which in turn is fully controlled by bosonic degrees of freedom in the thermodynamic limit. The second feature of our path-integral approach is more technical one because it allows to simplify the further calculations of the Bose polaron properties in terms of inverse powers of $N$.

The impurity single-particle Green's function $\mathcal{G}_f(P)=\langle \psi^*_{P} \psi_{P} \rangle$, where $\langle\ldots \rangle$ denotes the statistical averaging with action (\ref{A_eff}),
is uniquely determined by the self-energy part $\Sigma_f(P)$ 
\begin{eqnarray}
	\mathcal{G}^{-1}_f(P)=i\nu_p-\xi_f(p)-\Sigma_f(P),
\end{eqnarray}
\begin{figure}
	\centerline{\includegraphics
		[width=0.35\textwidth,clip,angle=-0]{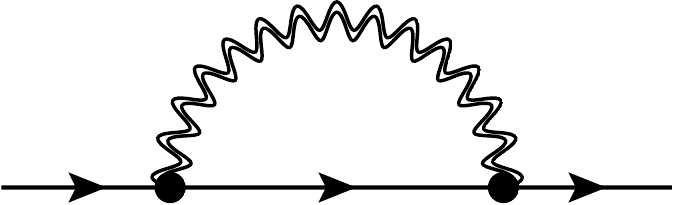}}
	\caption{The impurity self-energy in the large-$N$ limit. The solid lines with arrows denote the zero-order impurity Green's functions; black dots stand for bare vertices $i\frac{\tilde{g}}{g}$. The double wavy line is the induced boson-fermion effective interaction presented in Fig.~2.}
\end{figure}
\begin{figure}
	\centerline{\includegraphics
		[width=0.7\textwidth,clip,angle=-0]{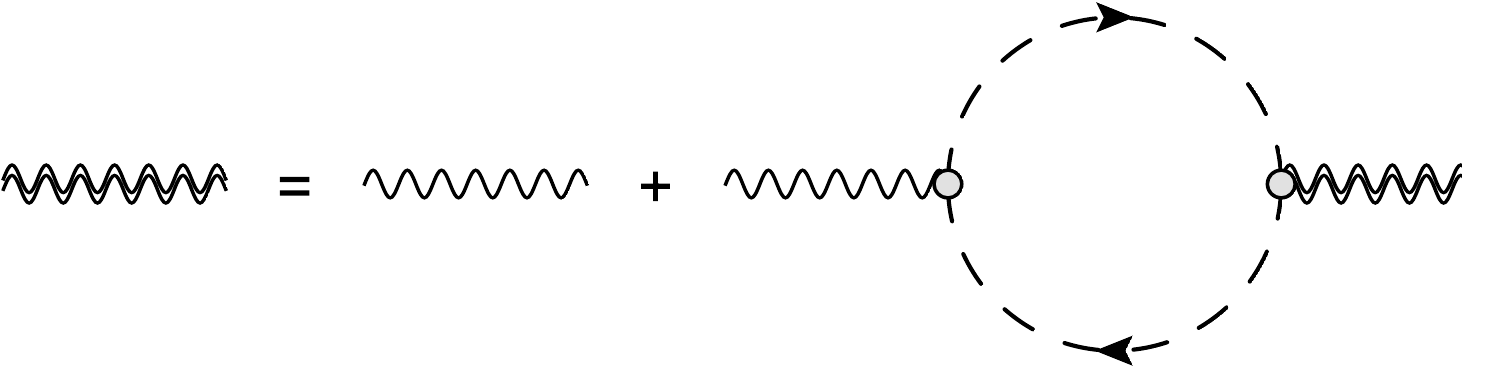}}
	\caption{Boson-mediated effective two-body impurity potential. The dashed line is the Green function of ideal Bose gas, wavy line and white dot denote $\frac{N}{g}$ and vertex $i$, respectively.}
\end{figure}
here the notation $\xi_f(p)=\varepsilon_f(p)-\tilde{\mu}_f$ is used.
In the large-$N$ limit singling out the Fock term (which is equal to zero in the one-fermion limit) one arrives at the self energy (see Figs.~1,2)
\begin{eqnarray}\label{Sigma_N}
	\Sigma_f(P)=\frac{T}{NV}\sum_K\frac{\tilde{g}^2 S(K)}{i(\omega_k+\nu_p)-\xi_f(|{\bf k}+{\bf p}|)},
\end{eqnarray}
with $S(K)=\Pi(K)/[1+g\Pi(K)]$ being the density-density correlation function for each sort of bosons. Close to the phase-transition point the bosonic polarization bubble reads
\begin{eqnarray}\label{Pi}
	\Pi(K)=\frac{T}{V}\sum_Q\frac{1}{i\omega_q-\varepsilon_q}\frac{1}{i(\omega_q+\omega_k)-\varepsilon_{|{\bf q}+{\bf k}|}}.
\end{eqnarray}
The evaluation of the leading-order self-energy correction (\ref{Sigma_N}) is rather simple, but nevertheless requires precision during the calculation. Particularly the summation over the Matsubara frequencies in this equation can be performed by using spectral theorem for the density-density correlation function
\begin{eqnarray}
	S(K)=\int_{-\infty}^{\infty}\frac{d\omega}{\pi}\frac{S_I(\omega,k)}{\omega-i\omega_k},
\end{eqnarray}
(here $S_I(\omega,k)={\rm Im} S(K)|_{i\omega_k\to \omega+i0}$) and gives for the self energy in the one-fermion limit ($\mu_f\to -\infty$)
\begin{eqnarray}\label{Sigma_omega}
	\Sigma_f(P)=\frac{\tilde{g}^2}{NV}\sum_{{\bf k}}\int_{-\infty}^{\infty}\frac{d\omega}{\pi} \frac{S_I(\omega,k)n(\omega/T)}{\omega+ i\nu_p-\xi_f(|{\bf k}+{\bf p}|)},
\end{eqnarray}
where $n(x)=1/(e^x-1)$ is the Bose distribution. It should be noted that for a Bose gas this $1/N$-result is equivalent to the well-known Random Phase Approximation (see, for instance \cite{Watabe}), which in turn leads to the Bogoliubov theory at low $T$. Despite its simplicity this approximation incorporates many important features of the finite-temperature properties of Bose systems, namely, the presence of second-order phase transition, universal power-law behavior of thermodynamic and structure functions, etc. Performing an analytical continuation of $\Sigma_f(P)|_{i\nu_p\to \nu+i0}=\Sigma^R_f(\nu,p)+i\Sigma^I_f(\nu,p)$ in the upper complex half-plane one obtains the $1/N$-corrected polaron energy
\begin{eqnarray}\label{vareps*}
	\varepsilon^*_f(p)=\varepsilon_f(p)+n\tilde{g}+\Sigma^R_f(\xi_f(p),p),
\end{eqnarray}
and damping
\begin{eqnarray}\label{Gamma*}
	\Gamma_f(p)=-\Sigma^I_f(\xi_f(p),p).
\end{eqnarray}
The structure of Green's function near the singularity is determined by the quasiparticle residue calculated with the same accuracy
\begin{eqnarray}\label{Z}
	Z^{-1}_f(p)=1-\frac{\partial}{\partial \xi_f(p)}\Sigma^R_f(\xi_f(p),p).
\end{eqnarray}

At zero temperature the result (\ref{Sigma_omega}) reproduces to the first-order beyond mean-field correction for the self energy \cite{Grusdt_Demler} obtained within the Rayleigh-Schr\"odinger perturbation theory. The $1/N$-expansion alters non-trivially the finite temperature region only. For a dilute Bose gas at non-zero temperatures the physics of the system is dictated by the infrared behavior of functions standing under the integral in Eq.~(\ref{Sigma_N}), therefore one may replace $n(x)$ by its asymptote $1/x$. This observation allowed \cite{Hryhorchak} to predict qualitatively correct behavior \cite{Baym_et_al} of the BEC transition temperature for a weakly-interacting Bose gas. Then the integration over $\omega$ in formula (\ref{Sigma_omega}) yields for the energy correction
\begin{eqnarray}\label{Sigma_as}
	&\Sigma^R_f(\xi_f(p),p)=\frac{\tilde{g}^2T}{NV}\sum_{{\bf k}}\frac{1}{\varepsilon_f(|{\bf k}+{\bf p}|)-\varepsilon_f(p)}\nonumber\\
	&\times\left[S_R(\varepsilon_f(|{\bf k}+{\bf p}|)-\varepsilon_f(p),k)-S_R(0,k)\right],
\end{eqnarray}
where only infrared asymptote of the real part $S_R(\omega,k)={\rm Re} S(K)|_{i\omega_k\to \omega+i0}$ of the retarded density-density correlator should be taken into account. Thus, the problem is reduced to the calculation of the above integral and function $S_R(\omega,k)$ (see Appendix A).

\section{Results}
A little unexpected thing about result (\ref{Sigma_as}) is that for a dilute Bose bath ($an^{1/D}\to 0$) the leading-order binding energy correction as well as the damping of a motionless (${\bf p}=0$) impurity are calculated analytically. The same conclusions can be drawn for the effective mass and for the quasiparticle residue. But the spatial dimensionality impacts crucially in the behavior of a Bose polaron therefore in the following we will consider $2<D<3$ and $D=3$ cases separately. 
\subsection{$2<D<3$}
It is well-known that for Bose systems in $D>2$ the condensation phenomenon occurs at finite temperatures. This second-order transition point is characterized by the developed density fluctuations which influence enormously on the impurity properties. Even for our model with a short-ranged two-body potential the leading-order induced boson-fermion effective interaction (see Fig.~2) has a power law behavior at large interparticle spacing. Although this large-$N$ approach effectively sums up infinite series of diagrams but it still possesses the perturbative nature in terms of the boson-impurity coupling parameter $\tilde{g}$. As a consequence we assume through out the text that the ratio $\tilde{a}/a$ is of order unity while $g\to 0$. The second restriction is more technical and is set on the mass of impurity atom: our leading-order calculations require the bare polaron mass to be not too close to the mass of surrounding Bose particles.  

Substituting formulae from the Appendix A in Eq.~(\ref{Sigma_as}) and calculating simple integral for the correction to energy of a motionless polaron we obtain
\begin{eqnarray}\label{Sigma_LO}
	\frac{\Sigma^R_f(-\tilde{\mu}_f,0)}{n\tilde{g}}=\frac{1}{N}\left(\frac{\tilde{a}}{a}\right)^{D-2}\left(an^{1/D}\right)^{\frac{(D-2)^2}{4-D}}\epsilon_D(\gamma),
\end{eqnarray}
where $\gamma=m/m_f$. The applicability of the obtained here and below formulas is not restricted to the integer dimension Bose polarons but can be easily used for impurities immersed in Bose condensates on fractals. Particularly in Fig.~3 the dimensionless function  $\epsilon_D(\gamma)$ is depicted for a fractional spatial dimension $D=\ln(20)/\ln(3)$ of the Menger sponge. It should be noted that the limits of applicability of the above result is restricted to $an^{1/D}\ll |1-\gamma|^{\frac{3-D}{D-2}}$. This limitation is related to the simplifications performed during the calculation of the integrals. Here we mean the replacement $n(x)\to 1/x$ and account of the leading-order infrared asymptote of the real and imaginary parts of the polarization operator. Actually, close to $\gamma =1$ we have to go beyond the approximation for functions $I(\omega,k)$ and $R(\omega,k)$ presented in the Appendix A, which will necessarily change the exponent in the power-law dependence on the gas parameter $an^{1/D}$ in Eq.~(\ref{Sigma_LO}). Exactly in the same manner we have calculated the leading-order damping (see Fig.~3)
\begin{eqnarray}
	\frac{\Gamma_f(0)}{n\tilde{g}}=\frac{1}{N}\left(\frac{\tilde{a}}{a}\right)^{D-2}\left(an^{1/D}\right)^{\frac{(D-2)^2}{4-D}}\Delta_D(\gamma).
\end{eqnarray}
\begin{figure}
	\centerline{\includegraphics
		[width=0.8\textwidth,clip,angle=-0]{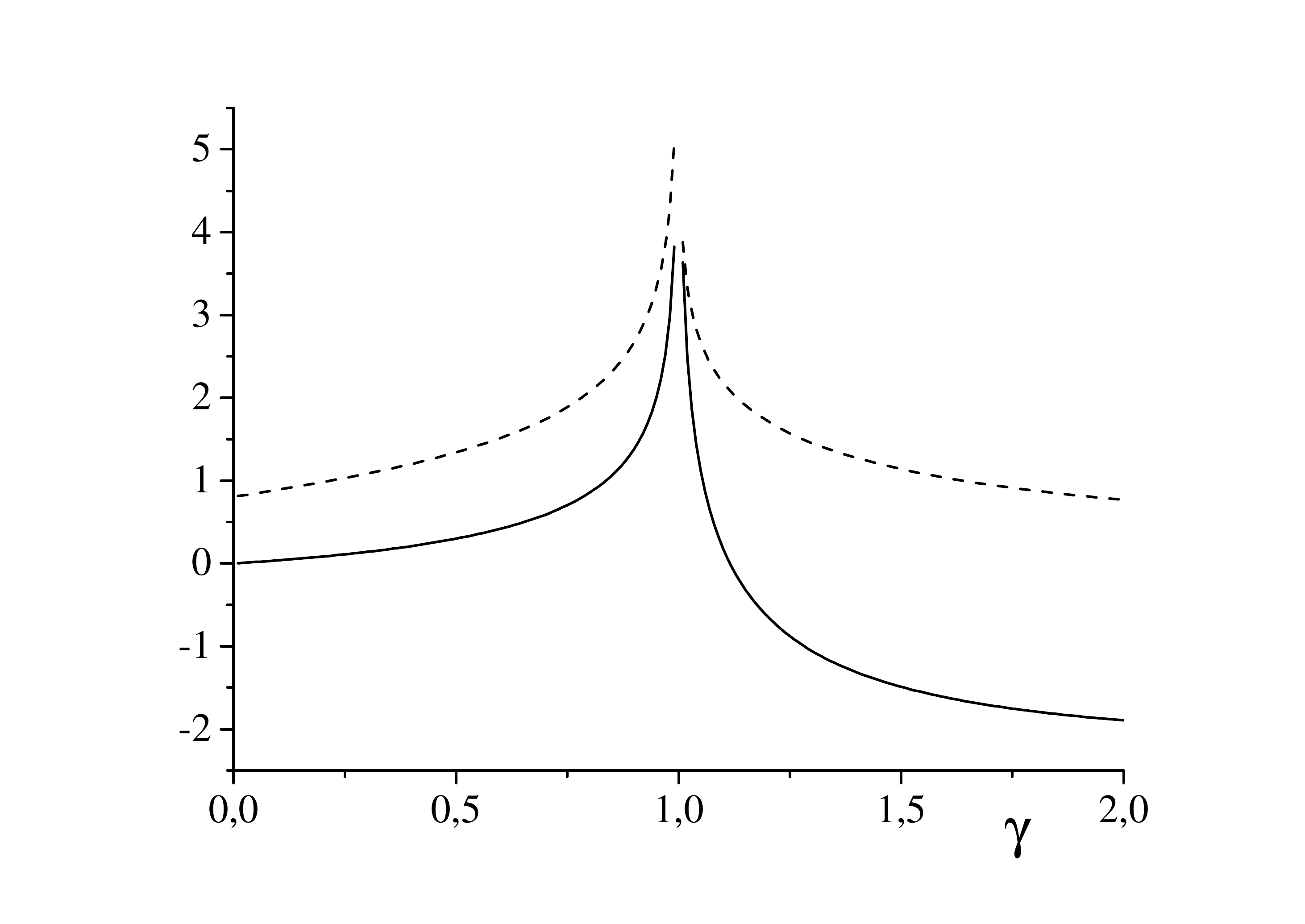}}
	\caption{Functions $\epsilon_D(\gamma)$ (solid line) and $\Delta_D(\gamma)$ (dashed line) determining the energy correction and the damping for a system with dimensionality of the Menger sponge $D=\ln(20)/\ln(3)$, respectively.}
\end{figure}
We recall again that our calculations are incorrect in the vicinity of the equal-mass limit and the exact damping to the order $1/N$ is equal to zero when $\gamma=1$.
Totally different situation is met in the finite-momenta dependence of the self energy.
In particular, trying to calculate the correction to the impurity effective mass by expanding $\Sigma^R_f(\xi_f(p),p)$ to quadratic order in $p$ we find out that the appropriate integral is logarithmically divergent in any dimension. This problem also appears during the calculation \cite{VPP_1} of effective mass of Bose particles in the critical point where these divergences are treated as signature of the power-law behavior  $p^{2-\eta_{\varepsilon}}$ of the one-particle spectrum. The critical exponent $\eta_{\varepsilon}$ in that case is universal, i.e., dependent on the global parameters only. If we assume qualitatively the same behavior for the polaron renormalised dispersion relation we immediately obtain the result valid for {\it any} $g$ in the adopted approximation 
\begin{eqnarray}\label{varepsilon*}
	\lim_{p\to 0}[\varepsilon^*_f(p)-\varepsilon^*_f(0)]\propto p^{2-\eta_{\varepsilon}},
\end{eqnarray}
with the non-universal exponent, which is given by 
\begin{eqnarray}
	\eta_{\varepsilon}=\frac{1}{N}\left(\frac{\tilde{a}}{a}\right)^{2(D-2)}\eta_{\varepsilon}(\gamma),
\end{eqnarray}
on the $1/N$-level and depicted in Fig.~4 for various mass ratios.
\begin{figure}
	\centerline{\includegraphics
		[width=0.8\textwidth,clip,angle=-0]{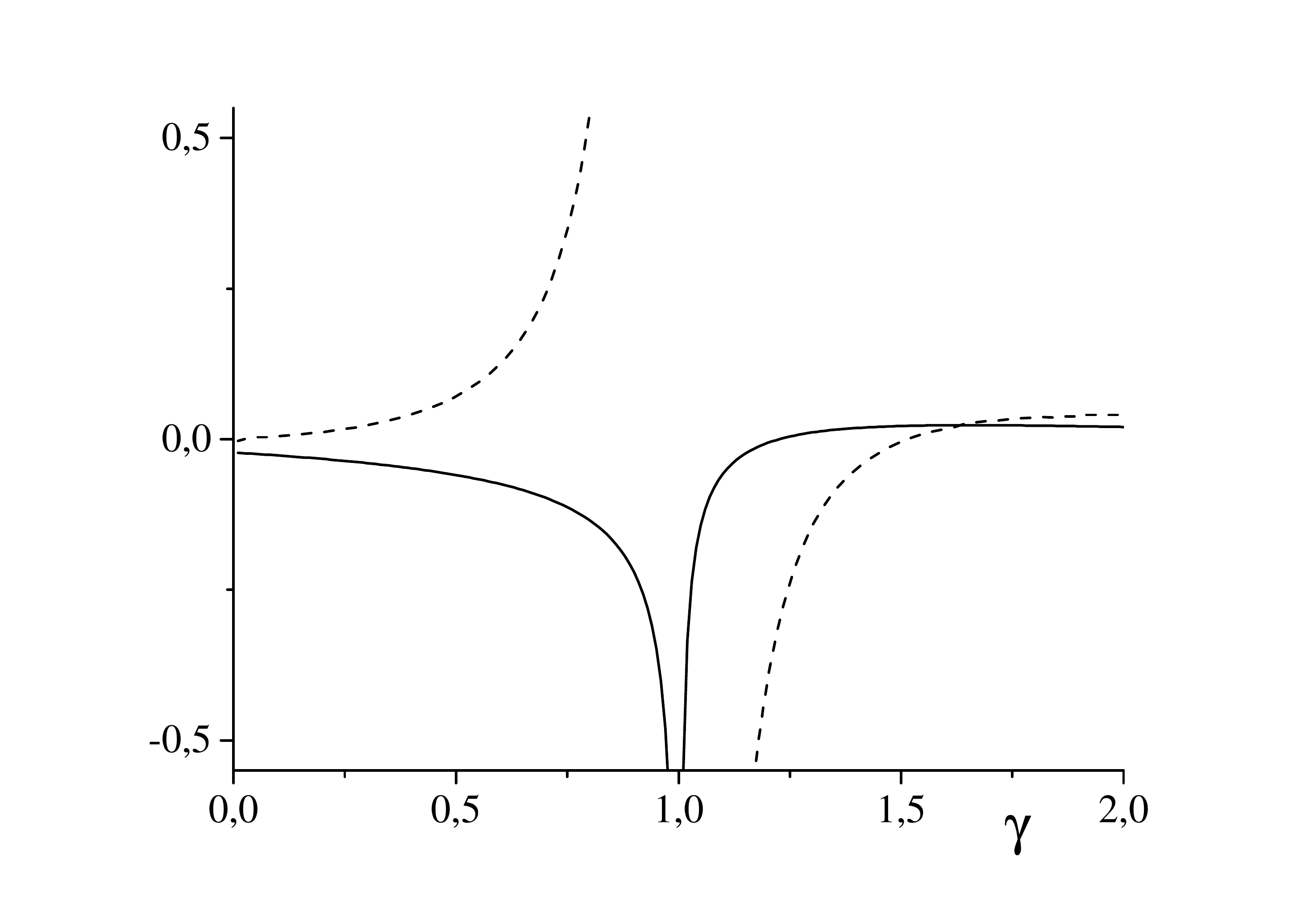}}
	\caption{Dimensionless functions $\eta_{z}(\gamma)$ (solid line) and $\eta_{\varepsilon}(\gamma)$ (dashed line) for $D=\ln(20)/\ln(3)$.}
\end{figure}
It is not surprising the quasiparticle residue (\ref{Z}) to demonstrate the non-analytic dependence on the wave-vector in the long-length limit too. This suggest the following behavior of the retarded impurity Green's function in the BEC point
\begin{eqnarray}\label{Gsing}
	\lim_{p \to 0}\mathcal{G}^R_f(\nu,p)|_{\nu\to \varepsilon^*_f(p)}\propto \frac{1}{[\nu-\varepsilon^*_f(p)]^{1+\eta_{z}}},
\end{eqnarray}
where exponent $\eta_{z}$ that differs from $\eta_{\varepsilon}$ and equals to
\begin{eqnarray}
	\eta_{z}=\frac{1}{N}\left(\frac{\tilde{a}}{a}\right)^{2(D-2)}\eta_{z}(\gamma).
\end{eqnarray}
The obtained curves (see Figs.~1,2) clearly indicate the difference in the behavior of heavy ($\gamma<1$) and light ($\gamma>1$) impurities when $2<D<3$.

\subsection{$D=3$}
This difference is even more drastic in the tree-dimensional case where the dependence on gas parameter $an^{1/3}$ of the leading-order correction to the energy of motionless polaron is different for $\gamma<1$ and for $\gamma>1$. Particularly for light impurity we have found (with logarithmic accuracy)
\begin{eqnarray}\label{Sigma_LO3}
	\frac{\Sigma^R_f(-\tilde{\mu}_f,0)}{n\tilde{g}}=\frac{1}{N}	\tilde{a}n^{1/3}\ln[an^{1/3}] \epsilon_3(\gamma),
\end{eqnarray}
($\gamma>1$). The second correction to the energy of heavy atom is given by Eq.~(\ref{Sigma_LO}) with function $\epsilon_3(\gamma)$ (see Appendix B)
\begin{eqnarray*}
	\epsilon_3(\gamma)=\frac{4\pi(1+1/\gamma)}{[\zeta(3/2)]^{4/3}}\Big\{\delta_3(\gamma)[\pi/2-\arctan(1/\delta_3(\gamma))]-\ln\sqrt{1+\delta^2_3(\gamma)}\Big\},
\end{eqnarray*}
where $\delta_3(\gamma)=i_3(\gamma)/r_3(\gamma)=\frac{1}{\pi}\ln\left|\frac{1+\gamma}{1-\gamma}\right|$.
The leading-order damping instead does not change its behavior thoroughly with the deviation of the mass ratio from $\gamma=1$ and equals to
\begin{eqnarray}\label{Gamma_LO3}
	\frac{\Gamma_f(0)}{n\tilde{g}}=-\frac{1}{N}\frac{4\pi(1+1/\gamma)\delta_3(\gamma)}{[\zeta(3/2)]^{4/3}}\tilde{a}n^{1/3}\ln[an^{1/3}].
\end{eqnarray}
In the three-dimensional case the obtained formulae due to the performed simplifications in the computation of integrals are valid only if $an^{1/3}\ll 1/\ln(1-\gamma)$. Similarly to the general consideration from previous section we have determined the critical exponents controlling behavior of the impurity Green's function. The peculiarity of the spectrum in the long-wavelength limit is contained in
\begin{eqnarray}
	\eta_{\varepsilon}(\gamma)=\frac{2}{3\pi^2}(1+\gamma)^2\gamma\frac{\partial^2}{\partial \gamma^2}\frac{1}{\gamma}\left\{\begin{array}{c}
		\frac{\delta^2_3(\gamma)}{1+\delta^2_3(\gamma)}, \ \ \gamma<1\\
		\ \ \ \ 1,   \ \ \ \ \ \ \gamma\ge 1
	\end{array}\right.,
\end{eqnarray}
and an exponent of the branch-point singularity of $\mathcal{G}^R_f(\nu,p)$ reads
\begin{eqnarray}
	\eta_{z}(\gamma)=-\frac{1}{2\pi^2}(1+\gamma)^2\frac{\partial}{\partial \gamma}\frac{1}{\gamma}\left\{\begin{array}{c}
		\frac{\delta^2_3(\gamma)}{1+\delta^2_3(\gamma)}, \ \ \gamma<1\\
		\ \ \ \ 1,   \ \ \ \ \ \ \gamma\ge 1
	\end{array}\right..
\end{eqnarray}
\begin{figure}
	\centerline{\includegraphics
		[width=0.8\textwidth,clip,angle=-0]{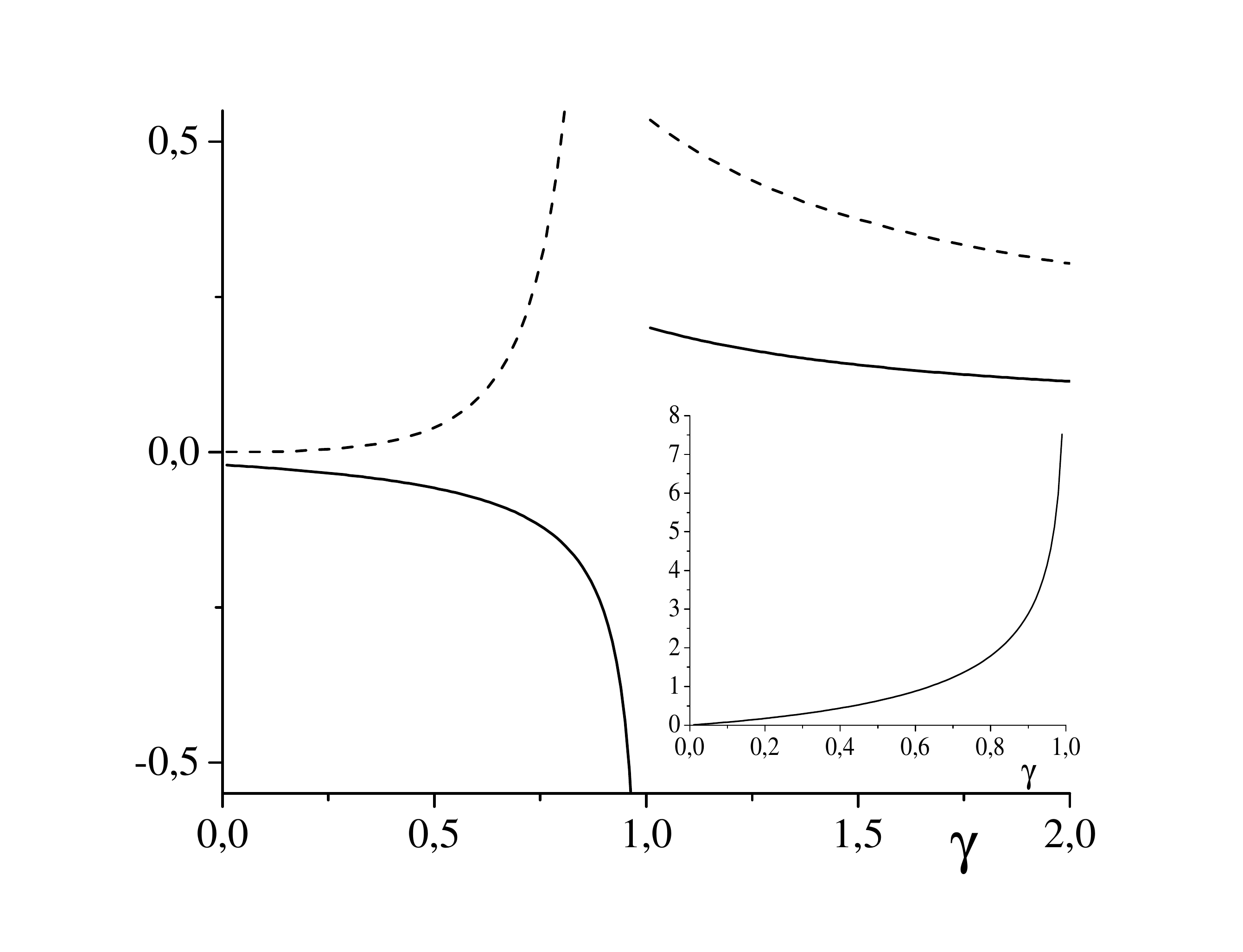}}
	\caption{Three-dimensional functions $\eta_{z}(\gamma)$ (solid line) and $\eta_{\varepsilon}(\gamma)$ (dashed line). The inset shows dimensionless energy $\epsilon_3(\gamma)$ of heavy three-dimensional impurity ($\gamma<1$).}
\end{figure}
The leading-order $1/N$-expansion is known to be insensitive to the many-body statistical effects. For instance, parameters of the impurity spectrum at low temperatures in the equal-mass limit and with $\tilde{a}$ replaced by $a$ reproduce the one-particle spectrum of medium bosons (compare the fist-order correction to effective mass in Ref.~\cite{Ardila_1} and the low-temperature (large-$g$) result for the one-particle spectrum in Ref.~\cite{Hryhorchak}). Interestingly that the same correspondence is observed in the critical temperature. Indeed, identifying impurity with the environmental particle ($\gamma\to 1+0$ and $\tilde{a}=a$) and calculating the zero-frequency Green's function at small $p$ with taking into account Eqs.~(\ref{varepsilon*}), (\ref{Gsing}) and imposing $\tilde{\mu}_f=0$ (recall that the inverse one-particle Green's function of bosons satisfies identity $\mathcal{G}^{-1}(0)=0$ at the BEC temperature) we immediately find $\lim_{p \to 0}\mathcal{G}^{-1}_f(P)|_{\nu_p = 0}\propto (p^{2-\eta_{\varepsilon}})^{1+\eta_z}=p^{2-\eta}$, where in the adopted approximation $\eta=\eta_{\varepsilon}-2\eta_z=\frac{16}{3\pi^2N}-2\frac{2}{\pi^2N}=\frac{4}{3\pi^2N}$, which is in agreement with the large-$N$ evaluations of the Fisher exponent \cite{Ferrell,Abe,Vakarchuk} in $D=3$.

\section{Conclusions}
Summarizing, we have considered the properties of a mobile impurity immersed in the $D$-dimensional superfluid across the BEC temperature. By using the large-$N$ expansion we have calculated the leading-order (beyond trivial Hartree term) correction to the polaron spectrum in the limit of a very dilute bosonic bath. It was shown that the results for the energy and damping strongly depend on the mass ratio of impurity atom and medium-forming particles. Our findings also revealed the infrared structure of the impurity Green's function that was found to demonstrate the branch point singularity in the critical region.

\section*{Acknowledgements}
We are grateful to Prof. I.~Vakarchuk and Prof. A.~Rovenchak for useful suggestions.
This work was partly supported by Project FF-30F (No.~0116U001539) from the Ministry of Education and Science of Ukraine.

\section{Appendices}
\subsection{Appendix A: Polarization operator}
The evaluation of the retarded density-density correlator requires the knowledge of analytically continued ``particle-hole'' bubble $\Pi(K)|_{i\omega_k\to \omega+i0}=R(\omega,k)+iI(\omega,k)$. After the Matsubara frequency summation in Eq.~(\ref{Pi}) the imaginary part $I(\omega,k)$ is easily obtained
\begin{eqnarray*}
	I(\omega,k)=\frac{\pi}{V}\sum_{{\bf q}}n(\varepsilon_q/T)[\delta(\varepsilon_{|{\bf q}+{\bf k}|}-\varepsilon_q-\omega)-(\omega\to -\omega)].
\end{eqnarray*}
Due to presence of the $\delta$-function the integration is relatively simple yielding at the critical temperature
\begin{eqnarray*}
	I(Tu,kk_0)=\frac{\sqrt{\pi}n}{2\zeta(D/2)Tk}\left[G_{\frac{D-1}{2}}(e^{-(k/2-u/2k)^2})-(u\to -u)\right],
\end{eqnarray*}
where $k_0=\sqrt{2mT}/\hbar=2\sqrt{\pi}[n/\zeta(D/2)]^{1/D}$ denotes the characteristic inverse length scale, $\zeta(x)$ is the Riemann zeta function and $G_{\alpha}(z)=\sum_{l\ge 1}z^l/l^{\alpha}$. In the infrared region ($u,k \to 0$), the leading-order terms in square brackets read
\begin{eqnarray*}
	\Gamma(3/2-D/2)\left[(2k)^{3-D}/|k^2-u|^{3-D}-(u\to -u)\right],
\end{eqnarray*}
($\Gamma(x)$ is the gamma function) and by applying the spectral representation $R(\omega,k)=P.V.\int_{-\infty}^{\infty}\frac{d \omega'}{\pi}\frac{I(\omega',k)}{\omega'-\omega}$
we obtained the real part of polarization operator with the same accuracy
\begin{eqnarray*}
	R(Tu,kk_0)=\frac{\sqrt{\pi}n}{2\zeta(D/2)Tk}\frac{\Gamma(3/2-D/2)}{\cot[(3/2-D/2)\pi]}\\
	\times\left[\frac{(2k)^{3-D}}{|k^2-u|^{3-D}}{\rm sign}(k^2-u)+(u\to -u)\right].
\end{eqnarray*}

\subsection{Appendix B: Parameters of the low-energy impurity Green's function}
In the same approximation as it was used for the evaluation of Eq.~(\ref{Sigma_as}) we calculated the imaginary part of self energy which deduces to the following integral
\begin{eqnarray*}
	\frac{\Gamma_f(0)}{n\tilde{g}}=\frac{1}{N}\frac{\Omega_D\lambda_D^{\frac{D-2}{4-D}}}{\pi^{D/2}\zeta(D/2)\gamma}\frac{\tilde{g}}{g}\int^{\infty}_{0}\frac{dkki_D(\gamma)}{[k^{4-D}+r_D(\gamma)]^2+i^2_D(\gamma)},
\end{eqnarray*}
while the real part is given by:
\begin{eqnarray*}
	&\Sigma^R_f(-\tilde{\mu}_f,0)/n\tilde{g}=\frac{1}{N}\frac{\Omega_D\lambda_D^{\frac{D-2}{4-D}}}{\pi^{D/2}\zeta(D/2)\gamma}\frac{\tilde{g}}{g}\int^{\infty}_{0}dkk^{D-3}\\
	&\times\left[\frac{r_D(\gamma)[k^{4-D}+r_D(\gamma)]+i^2_D(\gamma)}{[k^{4-D}+r_D(\gamma)]^2+i^2_D(\gamma)}-\frac{r_D(0)}{k^{4-D}+r_D(0)}\right],
\end{eqnarray*}
where for convenience parameter $\lambda_D=\frac{\sqrt{\pi}ng}{2\zeta(D/2)T}$ is introduced, $\Omega_D$ is the area of $D$-dimensional sphere with unit radius and shorthand notations for functions
\begin{eqnarray*}
	i_D(\gamma)=2^{3-D}\Gamma(3/2-D/2)\left[\frac{1}{|1-\gamma|^{3-D}}-(\gamma\to -\gamma)\right],
\end{eqnarray*}	
and
\begin{eqnarray*}
	r_D(\gamma)=\frac{2^{3-D}\Gamma(3/2-D/2)}{\cot[(3/2-D/2)\pi]}\left[\frac{{\rm sign}(1-\gamma)}{|1-\gamma|^{3-D}}+(\gamma\to -\gamma)\right],
\end{eqnarray*} 
explicitly related to $I(\omega,k)$ and $R(\omega,k)$ respectively are used. Note that the above integrals for arbitrary spatial dimension $D$ can be calculated exactly with the help of residue theorem. In the tree dimensions the integrals are divergent providing the logarithmic dependence on the parameter $\lambda_D$. Both the quasiparticle residue and effective mass by means of parameter differentiation can be expressed via a single integral, which is logarithmically divergent in the long-length limit at critical temperature. Actually this divergence determines exponents $\eta_{\varepsilon}$
\begin{eqnarray*}
	\eta_{\varepsilon}=-\frac{1}{N}\frac{4\Omega_D}{D\pi^{\frac{D+1}{2}}}\left(\frac{\tilde{g}}{g}\right)^2\gamma\frac{\partial^2}{\partial \gamma^2}\frac{1}{\gamma}\left\{\frac{r_D(\gamma)}{r^2_D(\gamma)+i^2_D(\gamma)}-\frac{1}{r_D(0)}\right\},
\end{eqnarray*}
and $\eta_{z}$ 
\begin{eqnarray*}
	\eta_{z}=\frac{1}{N}\frac{\Omega_D}{\pi^{\frac{D+1}{2}}}\left(\frac{\tilde{g}}{g}\right)^2\frac{\partial}{\partial \gamma}\frac{1}{\gamma}\left\{\frac{r_D(\gamma)}{r^2_D(\gamma)+i^2_D(\gamma)}-\frac{1}{r_D(0)}\right\}.
\end{eqnarray*}

\end{document}